# Challenges in addressing student difficulties with basics and change of basis for two-state quantum systems using a multiple-choice question sequence in online and in-person classes

Peter Hu*, Yangqiuting Li, and Chandralekha Singh

*Department of Physics and Astronomy, University of Pittsburgh, Pittsburgh, PA 15260*

*Corresponding author, e-mail: pth9@pitt.edu

**Abstract**

Research-validated multiple-choice questions comprise an easy-to-implement instructional tool for scaffolding student learning and providing formative assessment of students' knowledge. We present findings from the implementation of a research-validated multiple-choice question sequence on the basics of two-state quantum systems, including inner products, outer products, translation between Dirac notation and matrix representation in a particular basis, and change of basis. This study was conducted in an advanced undergraduate quantum mechanics course, in both online and in-person learning environments, across three years. For each cohort, students had their learning assessed after traditional lecture-based instruction in relevant concepts before engaging with the multiple-choice question sequence. Their performance was evaluated again afterwards with a similar assessment and compared to their earlier performance. We analyze, compare, and discuss the trends observed in the three implementations.

**Introduction**

Two-state systems are often used to illustrate many rich phenomena in quantum mechanics (QM), due to their relative simplicity compared to higher dimensional Hilbert spaces. Furthermore, since we are in the midst of the second quantum revolution [1,2], they are critical to the field of quantum information science to describe the behavior of qubits, the smallest unit in which quantum information is stored and processed. Yet because of the unfamiliarity with the quantum formalism, even advanced undergraduate students can often struggle with the requisite basic concepts, e.g., the distinction between inner and outer products, translation between Dirac notation and matrix representation and the concepts of basis and changing the basis to describe a quantum state. Prior research suggests that students in quantum mechanics courses often struggle with many common difficulties. Some prior studies have focused on topical investigations, with ideas such as quantum tunneling, relativistic quantum mechanics, and the Stern-Gerlach experiments [3–5]. Others have investigated broader issues in introductory through graduate-level quantum education, e.g., lost opportunities for engaging motivated students, visualization of quantum states in different cases, teaching quantum on an introductory level to get students to develop intuition, or investigation of graduate student understanding of various quantum concepts [6–9]. Yet others have provided extensive overviews of student difficulties [10–13] or focused on epistemological considerations in quantum instruction [14]. Some of the other areas of student struggle that have been investigated include the basic formalism [15,16], wavefunctions [17,18], measurement [15,17,19,20], and transferring learning from one context to other contexts [12,21]. Prior research also suggests that research-validated learning tools can effectively address such difficulties [22–25], including with concepts of basis and change of basis [26,27]. For example, Quantum Interactive Learning

Tutorials (QuILTs), developed and validated by our group, have been implemented with positive learning outcomes on a number of QM topics [28–30]. As another example, clicker questions, first popularized by Mazur [31] using his *Peer Instruction* method [31], have similarly been shown to be effective [32]. We build on this idea by developing and validating Clicker Question Sequences (CQS), which help students learn concepts using sequences of related questions [33–40]. Here, we describe the development, validation, and implementation of a CQS intended to help students learn about the basics of two-state quantum systems, including inner and outer products, translation between Dirac notation and matrix representation, and change of basis.

**Theoretical framework**

Available class time is a pointedly limited resource in efforts to engage students in the learning process and strengthen student learning outcomes. This is particularly true for QM courses, in which the content can often be counterintuitive and appear inconsistent with students' prior experiences in everyday life as well as other classical physics courses. Instructors, therefore, must consider research-based pedagogical approaches to engage students with these challenging concepts. With this background, our theoretical framework hinges on two different aspects of research-based pedagogical approaches, the balancing of innovation and efficiency [41] and taking advantage of peer interaction.

Schwartz et al.'s Preparation for Future Learning (PFL) framework [41] emphasizes the importance of both efficiency and innovation in instruction. The concepts of efficiency and innovation have been interpreted in different but related ways; e.g., efficiency can be used to refer to students' facility with widely applicable routine tasks that are retrieved frequently, such as mathematical manipulations that may be necessary but are not considered central to conceptual aspects of QM. On the other hand, innovation is the ability to apply existing knowledge to novel situations. Exploratory labs are examples of environments that are intended to maximize innovation. The authors observe that, in most traditional classrooms, efficiency is emphasized while innovation is typically disregarded, and suggest that balancing innovation and efficiency in learning activities can improve conceptual understanding and transfer of knowledge to new contexts [41].

In a separate publication, Schwartz and Martin used "efficiency" to describe aspects of several different instructional conditions, noting that invention tasks are significantly less efficient than more direct ways of delivering instruction, including lectures and readings, which are faster and more effective at teaching students to produce the correct answers. However, there can be a "time for telling" (or lecture) after students have struggled with innovative invention tasks, and they shed light on helpful effects that productive struggle during invention tasks can provide for students' learning [42]. Nokes-Malach and Mestre [43] analyzed Schwartz et al.'s study further using their own model of knowledge transfer and proposed that allowing students to struggle through invention tasks imparts a "hidden" effectiveness in several key ways. First, the particular framing and learning environment may prime students to operate in a more mastery-based orientation rather than a performance-based one; a mastery orientation is one in which students are interested in deeply understanding the material, while performance orientation implies students' desire to, e.g., get a good score or pass the course [44]. Similarly, the students in the condition resembling direct instruction may use problem-solving steps via a template without comprehensively considering the reasoning for each step. The authors concluded that, in that study, as a result of their deeper, mastery-oriented and highly-engaged struggle, students in the invention condition may have identified difficulties during their problem-solving session and remained unsatisfied with their initial solutions, which may have primed them to learn better subsequently. As a result, further instruction led them to outperform direct instruction students even when both groups were subsequently given access to a common resource explaining the concepts. Meanwhile, without explicitly knowing about such possible holes in their understanding, the direct instruction students would not have been actively primed to be looking to fill in such holes [43].

Having outlined the benefits of incorporating innovation, the aforementioned fixation on efficiency in most solely lecture-based classrooms leads to what Schwartz et al. term "routine experts," who are fluent in a specific type of task but struggle with transferring their learning to new situations, rather than "adaptive experts" who are able to complete a wide variety of tasks requiring transfer of knowledge in their area of expertise with speed and accuracy. A pure focus on innovation leads to another issue: the "frustrated novice," who without guidance and scaffolding support that efficient aspects of instructional design may provide is unable to make any meaningful progress on problems in a given amount of time. Thus, the authors conclude, developing students' expertise in a domain requires balancing of efficiency and innovation axes in the instructional design, outlining an "optimal adaptability corridor" by which they can develop competencies to become "adaptive experts" with the least amount of wasted effort [41].

Integrating a CQSs with lectures and using collaborative learning and whole-class discussions when incorporating research-based CQSs offer one mechanism for balancing efficiency and innovation in instructional design. This type of instructional design, in which CQS and lectures are integrated and students are provided opportunities to engage in discussions with peers and instructors, can help students develop competencies through the efficiency of lectures paired with the innovation of productive struggle during CQS administration and discussion. This approach can provide students opportunities to have small-group or whole-class discussions about specific concepts that may reinforce the benefits of lectures and deliberations. Moreover, when student difficulties are used as resources [45] to guide the development, validation, and evaluation of the CQS, prominent patterns of student thought can be efficiently addressed by letting students innovatively explore the concepts that need strengthening. Students typically reach a consensus of the correct answer about 80% of the time, and for the remaining cases, the instructor can clarify what the correct answer is.

Collaborative learning can be productive in physics classrooms [31,46], particularly when individual accountability and positive interdependence have been suitably incentivized. Methods to incentivize students include grade incentives and appropriate framing of the activities, such as by making clear to students that, if they strive to develop their communication skills and ability to work with others while in the low-stakes environment, they are setting foundations for future personal and professional growth. Under such conditions, there are several mechanisms by which collaborative learning may be beneficial. Firstly, although students may not be at the same point in their learning of a particular concept, they may still understand one another's difficulties better than the instructor can, since they have all learned the concept recently. This makes them likely to more effectively and efficiently clarify issues related to the difficulties and solve the problem together [47,48]. Secondly, explaining their thoughts to one another demands full engagement and analysis of their own thought processes in order to effectively communicate, so efforts to explicitly present their knowledge give clarity to the students doing the explaining, as well as those who initially articulated their difficulties during collaborative learning. The fine-grained parsing of each member's relevant knowledge may lead to more careful consideration of the problem and even yield insights that were not obvious beforehand (leading to co-construction of knowledge) [49].

Thirdly, collaborative learning often creates inherent expectations of accountability for each student, and it has been suggested that students are often more comfortable interacting with peers about their difficulties than with an authority figure [50]. Such interactions can include communicating their difficulties and clarifying their thought processes related to concepts in a domain. Moreover, through practice in collaboration, students not only can develop more comfort and competencies in their communication skills [51–55], but collaborative learning has also been found to positively affect students' self-efficacy and other motivational beliefs, which are associated with better student performance and persistence in STEM courses [56,57]. Additionally, in-class collaboration can forge connections between students that they can continue to use as resources outside of class, for example, when completing homework assignments.

Peer collaboration has been shown to be an effective method for students to learn in previous work for a variety of contexts [46,58], including in physics [59,60]. Students' performance on conceptual physics questions can receive a substantial boost from working in pairs compared to when only working individually, even if only one or neither student knew the correct answer for any given question. In particular, student pairs in which neither student initially provided a correct answer were able to reach the correct answer 30% of the time, a phenomenon known as co-construction. This effect of elevated performance even persisted when the students were assessed again individually, pointing to significant retention [60]. There is also some evidence that these effects may apply to quantum mechanics concepts as well [61].

Emphasizing the importance of collaborative learning, Chi et al. proposed the ICAP framework, in which there are four broad modes of learning: Interactive, Constructive, Active, and Passive (ICAP). The Passive mode comprises methods of traditional and direct instruction, such as through classroom lectures, readings, or videos, in which students only engage through visual and auditory channels without any further activity. The Active mode is used in this framework to describe situations in which students are making choices or physical manipulations in their learning materials, but without generating explicit new knowledge or connections to old knowledge, which is the condition that satisfies the criteria of the Constructive mode. Finally, the Interactive mode is characterized by co-construction in small groups, which is the scheme described above, and is associated with the largest improvements in student retention and performance as compared to the Passive mode. As the wording suggests, the Interactive mode is only possible when all constituent students are already situated and working together in the Constructive mode. Broadly speaking, research shows that each successive mode subsumes the behaviors and benefits of all the modes that rank beneath it, in the descending order I/C/A/P [62,63].

The clicker questions, first popularized for use in physics courses by Mazur using the *Peer Instruction* technique, are intended to be conceptual multiple-choice questions posed to the class to which students reach a consensus by discussing in small collaborative groups. Mazur's method detailed in *Peer Instruction* falls under the Interactive mode, and has been associated with better learning outcomes including performance and retention [31,32].

The CQS questions are designed to be an Interactive and Constructive activity, requiring students to think about conceptual knowledge and their implications in new and different ways collaboratively. When students engage in discussion about the CQS questions in small groups, they work under the Interactive learning mode within the ICAP framework. Although this is the preferred mode, in the research presented here, during the online implementation, constraints imposed by time and the affordances of the technology resulted in a largely absent groupwork component for the CQS. Instead, students were simply given the questions to think about, and they answered the questions via individual polling. Therefore, under the ICAP framework, we consider the students to be in the Constructive learning mode while engaging with the CQS content in the online year, and in the Interactive mode during the in-person years.

**Methods**

*Description of CQS*

A description of the CQS, consistent with its learning objectives, is provided here. CQS 1.1-1.4 help students identify properties of two-state spin systems and spin-1/2 systems in particular. CQS 2.1-2.6 help students achieve fluency in translating between Dirac notation and matrix representation and calculating inner and outer products. Finally, using the knowledge about bases and products from the preceding question sequences, CQS 3.1-3.5 focus on helping students change the basis of a quantum state through several approaches. In particular, the three methods discussed to help students be able to change basis were direct substitution (e.g., using $|\pm x\rangle = \frac{1}{\sqrt{2}}(|+z\rangle \pm |-z\rangle)$ to replace a state expressed in the *x*-basis using the standard notation); the viewing of inner products, e.g., $\langle \pm x|\chi\rangle$ as projections of the state $|\chi\rangle$ along the basis vectors in a particular basis (in this case, the *x*-basis); and the use of spectral decomposition of unity. The clicker questions are provided in Appendix A, and the concepts on which they focus are summarized in Table 1.

*Development and validation*

The CQS on the basics of two-state quantum systems is intended for use in upper-level undergraduate QM courses. During the development and validation process, we took inspiration from a QuILT on Dirac notation [64], which contained several key concepts applicable to both the basics of two-state systems and methods of changing basis. This enabled us to build on previous work without starting over completely, since much of the work in cognitive task analysis, from both student and expert perspectives, had already been completed. The QuILT focuses on the contexts of both wavefunctions in an infinite-dimensional vector space and two-state systems, while the CQS adapts the content fully to two-state systems, with an eye toward condensing the material and highlighting the most important information in light of the greater time constraints. Using the insights gained from the development of the QuILT, including student interviews and investigations in authentic classroom environments, we adapted the relevant learning objectives and questions while supplementing them with new ones. This iterative process involved the input of researchers and other physics faculty members, incorporating many perspectives to ensure maximal clarity and consistency in the wording and framing of the questions. The CQS is designed to provide students opportunities to think about common difficulties, struggle productively, and get immediate feedback from their peers and instructors. The CQS also deliberately includes concrete questions, which provide opportunities for students to apply their knowledge in specific contexts, and abstract questions, which can help students generalize their understanding of the concepts and transfer their knowledge across contexts.

In this CQS, the questions are carefully sequenced to build on one another. For example, the same concept may be applied in different contexts or different concepts may be applied in similar contexts in two consecutive questions. Thus, students can compare and contrast the premise of consecutive questions to solidify their understanding of the concepts and build their knowledge structure. To facilitate the class discussion using peer instruction as mentioned earlier, we also added some slides prompting instructor-led discussion between some questions in the CQS, which can be used to review and emphasize the important concepts in the previous questions or discuss broader themes related to those questions.

After the initial development of the CQS, starting with the learning objectives adapted from the inquiry-based guided sequences in the QuILT as well as empirical data from student responses to existing individually-validated questions in previous years, we further validated the CQS by conducting individual interviews with five students in which they completed the pre-test, entire CQS, and post-test (the pre- and post-test are described in the "*Course implementation*" section below) using a think-aloud protocol. In these interviews, we asked students to think aloud while answering the questions to understand their reasoning, refraining from disturbing them so as not to disrupt their thought processes. After each question, we first asked students for clarification of the points they may not have made; we then led discussions with them on each choice as appropriate. The feedback from students helped in fine-tuning and refining the new questions, as well as ensuring that they were appropriately integrated with existing ones to construct an effective sequence of questions.

*Course implementation*

The data presented here are from administration of the validated CQS in a mandatory first-semester junior-/senior-level QM course at a large research university in the United States. The final version of the CQS was implemented in three consecutive years, one online and two in person, with some minor adjustments made between years to streamline the presentation of the concepts. The instructor who taught the online class also taught the second of the in-person classes discussed below, which enables us to draw comparisons between different classroom environments (online versus in-person) as well as different instructors.

During the online implementation via the conferencing software Zoom, the CQS was presented as a Zoom poll while the instructor displayed the questions via the "Share Screen" function. For the in-person implementations, the poll was replaced by a functionally similar classroom clicker system, and students were asked to think before discussing their responses with each other before answering each question via clickers. For each question, the instructor displayed the results after all students had voted, before a full class discussion of the validity of the options provided. The instructor made a significant effort to implement peer instruction using Zoom's Breakout Rooms feature, but it was unclear whether the benefits of splitting the class into these smaller virtual groups outweighed the substantial time loss. In addition to taking much longer to separate and reconvene than would be necessary in person, the online environment itself has norms and barriers that inhibit free and spontaneous discussion among individuals. Because of these difficulties, the peer instruction feature was not considered a major factor in the online administration, but was realized for the in-person administrations.

To determine the effectiveness of the CQS, we developed and validated a pre- and post-test containing questions on topics covered in the CQS. The post-test was a slightly modified version of the pre-test, containing changes such as use of different quantum states, but otherwise maintaining underlying conceptual similarity. The correspondence of concepts on the pre- and post-test to concepts discussed in the CQS is provided in Table 1. In both online and in-person classes, students completed the pre-test immediately following traditional lecture-based instruction on the topic. (Question 3 was added for the two in-person years, and did not appear in the online year.) Since all relevant material has been covered via lecture-based instruction before students engage with the CQS, the CQS constitutes a type of content review. After administration of the CQS over two to three class sessions, students completed the post-test. Two researchers graded half of the pre- and post-tests and, after discussion, converged on a rubric for which the inter-rater reliability was greater than 95%. Afterward, one researcher graded the remaining half of the assessments.

The pre- and post-test questions are reproduced in Appendix B. Correct answers are provided in bold. A detailed breakdown of student performance on the tested concepts is provided in the next section. In the closing sections, we compare the online implementation with both an in-person implementation with the same instructor as well as one with a different instructor. We also compare the two in-person implementations with each other to determine the generalizability of the CQS's usefulness.

**Table 1.** Summary of the concepts that were covered in the CQS, listed along with the pre-test/post-test questions and CQS questions that address them.

| Concept | Pre-/post-test question | Corresponding CQS questions |
|---|---|---|
| Basic concepts related to basis and Hilbert space | 1 | 1.1, 1.2 |
| Operators are diagonal in certain bases which are eigenstates of the operators | 6, 7 | 1.3, 1.4 |
| Expression for $|\pm x\rangle$ states in terms of $|\pm z\rangle$ states | 6, 7 | 2.1 |
| Inner products are scalars | 4 | 2.2, 2.6 |
| Outer products are operators (matrices) | 5 | 2.3, 2.5, 2.6 |
| Translation between Dirac notation and matrix representation in the $z$-basis | 2, 3 | 2.2, 2.3, 2.4, 2.5, 2.6 |
| A bra state in matrix representation is the transposed complex conjugate of its corresponding ket state | 4, 5 | 2.4 |
| Interpretation of expansion coefficients in a particular basis as projections of the ket state along the basis vectors | 6, 7 | 3.1 |
| Use of substitution (mathematical relation between different basis states) to express a state in another basis | 6, 7 | 3.2 |
| Use of the spectral decomposition of the identity operator to change basis | 6, 7 | 3.3, 3.4, 3.5 |

**Results and discussion**

Students overall did very well on the post-test following CQS instruction. Some questions were also consistently easy for students on the pre-test, after traditional lecture instruction, but there are also questions on which students notably struggled.

**Table 2.** Results of the online administration of the CQS via Zoom (online class). Comparison of pre- and post-test scores, along with normalized gain [65] and effect size as measured by Cohen's $d$ [66], for students who engaged with the CQS ($N = 29$). (Note: Question 3 was not asked in this year.)

| Question # | Pre-test mean | Post-test mean | Normalized gain | Effect size |
|---|---|---|---|---|
| 1 | 83% | 99% | 0.93 | 1.03 |
| 2 | 91% | 95% | 0.40 | 0.14 |
| 4 | 86% | 86% | -- | -- |
| 5 | 47% | 60% | 0.26 | 0.31 |
| 6 | 69% | 84% | 0.50 | 0.37 |
| 7 | 50% | 86% | 0.72 | 0.86 |

**Table 3.** Results of the first in-person administration of the CQS (in-person class 1). Comparison of pre- and post-test scores, along with normalized gain and effect size as measured by Cohen's $d$, for students who engaged with the CQS ($N = 25$).

| Question # | Pre-test mean | Post-test mean | Normalized gain | Effect size |
|---|---|---|---|---|
| 1 | 95% | 97% | 0.50 | 0.24 |
| 2 | 78% | 88% | 0.45 | 0.27 |
| 3 | 90% | 96% | 0.60 | 0.24 |
| 4 | 76% | 94% | 0.75 | 0.69 |
| 5 | 66% | 94% | 0.82 | 0.91 |
| 6 | 92% | 98% | 0.75 | 0.29 |
| 7 | 62% | 98% | 0.95 | 1.08 |

**Table 4.** Results of the second in-person administration of the CQS (in-person class 2). Comparison of pre- and post-test scores, along with normalized gain and effect size as measured by Cohen's $d$, for students who engaged with the CQS ($N = 27$).

| Question # | Pre-test mean | Post-test mean | Normalized gain | Effect size |
|---|---|---|---|---|
| 1 | 86% | 99% | 0.91 | 0.87 |
| 2 | 93% | 96% | 0.50 | 0.23 |
| 3 | 94% | 100% | 1.00 | 0.37 |
| 4 | 70% | 81% | 0.38 | 0.31 |
| 5 | 59% | 85% | 0.64 | 0.69 |
| 6 | 61% | 87% | 0.67 | 0.62 |
| 7 | 48% | 78% | 0.57 | 0.65 |

In light of the recommended interpretations of Cohen's $d$ effect sizes as small ~ 0.20, medium ~ 0.50, and large ~ 0.80, while also noting that both normalized gain and effect size are less meaningful when students perform well on the pre-test, we make the following observations.

We observe that students, in general, were quite proficient with questions 1-3 across all three years after traditional instruction, with small differences (e.g., in-person class 1 on question 2, in which the pre-test score was 78%; see Tables 2-4). Question 1 is about basics of Hilbert spaces and matrix representations, and their correspondence to physical observables. Question 2 asks students to change from Dirac notation to matrix representation, and question 3 was added after the first online year to investigate whether students were comfortable changing from matrix representation to Dirac notation. Reassuringly, students performed well on question 3 on pre-test after traditional instruction. In particular, on these questions, students scored very high on the pre-test, and they almost universally gave correct responses on the post-test, a promising sign that advanced undergraduate students are comfortable with these representations and translating between them.

Question 4 asks students to calculate an inner product of two given states. Students did not have very much difficulty with this on either the pre-test or the post-test, and many correctly identified that the answer was a scalar. The most common difficulty was not taking the complex conjugate when transforming one of the ket states to its corresponding bra. Additionally, a small number of students gave vectors or matrices as answers instead of scalars. This was mostly limited to the pre-test and rectified on the post-test. This could indicate that these students needed to be refamiliarized with the rules of matrix multiplication, as some students who had difficulty with this question did not write the row and column vectors in the correct order when multiplying. These attempts resulted in such nonstandard matrix multiplication as, for example, $(a \ \ b)\begin{pmatrix}c\\d\end{pmatrix} = \begin{pmatrix}ac & bc\\ad & bd\end{pmatrix}$ or $\begin{pmatrix}a\\b\end{pmatrix}(c \ \ d) = ab + cd$.

Question 5 asks students to calculate the outer product of two given states, which are the same states as in question 4. Students' most common mistake was providing the same answer as for question 4, a scalar rather than a matrix, for which zero credit was given. This was observed on many students' pre-test responses in all classes, and some students' post-test responses during the online class. Students in the in-person years performed better on the post-test, with some even writing their answers in Dirac notation, reinforcing that these students possessed a fluency between the Dirac notation and matrix representation. (The majority of students preferred using matrix representation, likely because the question had the given states in matrix representation, or because it offers more compact notation.) Those students who neglected to take the complex conjugate when finding the corresponding bra state did so for both questions 4 and 5. Additionally, some students found the bra state corresponding to the ket state other than the one indicated, but they were given full credit if they otherwise performed the inner product correctly. Among incorrect responses, it was also common for students to provide the transpose of the correct matrix. On a related note, some provided the correct answer but continued to show the nonstandard method for matrix multiplication described above, indicating that they knew the result of such an operation but had not mastered the mechanics of matrix multiplication. It is possible that those students who unintentionally transposed the matrix did not have a functional understanding of the rules of matrix multiplication.

Overall, judging by the pre-test scores, inner products appear to be easier to grasp than outer products, which is reasonable partly because they learn about scalar products but not outer products in introductory courses. Many students simply repeated the inner product calculation when asked to find each one. However, particular emphasis on this point in the CQS was enough to help greatly reduce this difficulty.

For questions 6 and 7, students were asked to change basis. Question 6 provided a state in the $x$-basis to change to the $z$-basis, and question 7 went in the reverse direction, going from the $z$-basis to the $x$-basis. These questions were deliberately left open-ended so that students could use the method that they were most comfortable with, as the CQS went over several distinct approaches to changing basis. Most students chose to substitute $|\pm x\rangle$ states with their expressions in the $z$-basis, $\frac{1}{\sqrt{2}}\big(|+z\rangle \pm |-z\rangle\big)$, which were given in the pre-test and post-test. In question 7, the reverse relationships $|\pm z\rangle = \frac{1}{\sqrt{2}}\big(|+x\rangle \pm |-x\rangle\big)$ were not explicitly provided. Since it is possible to add and subtract the $|\pm x\rangle = \frac{1}{\sqrt{2}}\big(|+z\rangle \pm |-z\rangle\big)$ equations to find the explicit relationships, direct substitution is still a viable method. Some students went through this algebra, while others correctly recognized that the relationship between the $x$-basis and $z$-basis is symmetrical, which was another valid justification. These extra steps, however, can be challenging for some students and cause cognitive overload as they try to process all of the information, so question 7 was considered to be more difficult than question 6. For example, while a considerable number of students were unable to make meaningful progress on these questions (or left them blank) on the pre-test, that number was higher for question 7 than question 6 in each class. Fewer students used the spectral decomposition of the identity operator or explicitly described their work in terms of taking projections along the new basis, which could point to a lesser emphasis on these methods in class and on homework assignments.

The most common mistake after traditional lecture-based instruction on the pre-test was to simply divide the expansion coefficients in the starting basis by $\sqrt{2}$. Students with this type of difficulty did not recognize that such a state is not normalized. While the details of how students arrived at this result varied between students, in many cases, it may stem from them discarding some of the inner products $\langle +x|+z\rangle, \langle +x|-z\rangle, \langle -x|+z\rangle$, and $\langle -x|-z\rangle$ in their attempts to obtain the final answer, resulting in an incomplete projection along the new basis. Some other students arrived at the conclusion that the expansion coefficients do not change when transforming from one basis to the other (i.e., $|\chi\rangle = a|+x\rangle + b|-x\rangle = a|+z\rangle + b|-z\rangle$), and this was most common among students for question 7 after correctly answering question 6. These notions were largely corrected on the post-test.

Additionally on the post-test, rather than finding $a|+x\rangle + b|-x\rangle$ in the $z$-basis as asked, one student was observed to instead find the state $b|+x\rangle - a|-x\rangle$, which is orthonormal to the given state. This was an interesting response, as it demonstrated an understanding of what makes an orthonormal basis, and possibly the idea that one can arbitrarily construct an infinite number of bases, though it had little to do with transforming from the given basis to the target basis. It is likely that this student had simply misinterpreted the question.

Overall, students who engaged with the CQS demonstrated performance on the post-test equal to or higher than that on the pre-test for every concept tested, and this held true across both instructors and both modes of instruction. This shows that the CQS helped students in grasping the concepts, substantially so for the more difficult concepts. Students' common difficulties that were addressed by the CQS are summarized in Table 5.

**Comparisons between in-person and online instruction**

During the online year, post-test performance on question 5, involving the calculation of an outer product, did not improve appreciably compared to some of the other questions on which students struggled. It is unclear why many students, even on the post-test, simply reiterated their answers for question 4, thus providing a scalar rather than a matrix. (Students in this online year also did not improve on question 4, regarding the inner product, but it is worth noting that the pre-test score was the highest of the three years; see Tables 2-4.) This was the only year in which this remained a widespread difficulty even after CQS instruction. This is somewhat puzzling, as performance on all the other questions in the online class is comparable to that in the second in-person class, which was taught by the same instructor. On the one hand, this could have been a concept that did not become as clear to students even after the CQS, thanks to the difficulties posed by the online environment. On the other hand, their methods of changing basis were sound in questions 6 and 7, showing that a larger increase in performance is possible when using the CQS in an online setting (see Table 2). It is possible that the focus on the basis change content in the CQS was enough for students to have their difficulties reduced even in the online administration, while the content on distinguishing the inner product from the outer product did not get through to the students quite as effectively. It could also be the case that the higher quality of class discussions during the in-person administrations helped students better understand outer products.

Aside from this, student scores from online instruction are seen to be only slightly worse or no worse than those from in-person instruction. Similarly, we have seen student performance in online classes in other studies to be on par in some respects [38,39]. However, students in the online class also all had their cameras and microphones off, possibly enabling them to consult resources that they were not intended to access, so there may also be some inflation in the scores, as was suggested in another study of student performance on content surveys [67].

In-person class 1 had the best performance out of all three years. The biggest improvements were observed in questions 5 and 7. In this year, the CQS on basics of two-state systems was administered after another CQS on the basics of Dirac notation, rather than serving as the very first CQS in the course as was the case in the other years. It is possible that students were more well-adjusted to learning from a CQS having already done the basics of Dirac notation CQS, compared to the online class and in-person class 2.

However, in-person class 1 did have more students than others show difficulty on question 2 on the pre-test after traditional instruction. While many students were able to correctly express certain states using matrix multiplication in the $z$-basis, a not insubstantial number gave responses along the lines of $\frac{1}{\sqrt{2}}\big(|+z\rangle - |-z\rangle\big) \doteq \frac{1}{\sqrt{2}}\begin{pmatrix} 1 & 0 \\ 0 & -1 \end{pmatrix}$. This difficulty was largely corrected on the post-test, and was not observed as a common response in other years.

**Comparisons between in-person classes**

On the whole, in-person class 1 did better of the two in-person classes on the pre-test. For example, in-person class 1 students, on average, struggled only on questions 5 and 7, while in-person class 2 students struggled on four questions (questions 4-7). However, on average, both in-person classes performed well on all questions on the post-test, suggesting that the CQS was effective in helping students in both classes. With regard to the differences, particularly in the pre-test scores, instructor-level and student-level fluctuations such as these are to be expected with different lecturing styles and course decisions. Moreover, both instructors taught with a spins-first approach, but they used different textbooks (McIntyre and a modified approach using Griffiths).

The question that students in in-person class 1 missed at higher rates than in either of the other classes was question 2, translating two ket states $|+y\rangle$ (for which example responses follow) and $|-x\rangle$ from Dirac notation to matrix representation. Aside from one blank answer, the incorrect responses provided $2 \times 2$ matrices, e.g., $\frac{1}{\sqrt{2}}\begin{pmatrix} 1 & 0 \\ 0 & i \end{pmatrix}$, rather than column vectors, e.g., $\frac{1}{\sqrt{2}}\begin{pmatrix} 1 \\ i \end{pmatrix}$. These types of incorrect responses were observed in other years as well, in addition to other types, which most often took the form of some mathematical oversight, e.g., $\begin{pmatrix} 1 \\ \frac{i}{\sqrt{2}} \end{pmatrix}$ or $\frac{1}{\sqrt{2}}\begin{pmatrix} 1+i \\ i \end{pmatrix}$. In addition to natural variance, instructor- and student-level differences may have played a part.

However, incorrect responses were very much the exception rather than the norm in all classes (see Tables 2-4).

Despite the differences, both classes have clearly benefited from the CQS, just as students benefited in both modalities.

**Table 5.** Student difficulties on the pre-test, successfully addressed by CQS instruction and reduced on the post-test.

| Difficulties | CQS # | Pre-/post-test # \| comments |
|---|---|---|
| Basics of basis and Hilbert space | 1.2 | 1 \| Some improvement, high pre-test scores |
| A bra state in matrix representation is the transposed complex conjugate of its corresponding ket state | 2.4 | 4, 5 \| Some improvement online, major improvement in-person |
| Outer products in two-dimensional Hilbert space are 2 × 2 matrices, not scalars | 2.3, 2.5, 2.6 | 5 \| Some improvement online, major improvement in-person |
| Changing from $x$-basis to $z$-basis (possible with substitution method, since the $x$-basis states were provided in terms of $z$-basis states) | 3.2 | 6 \| Some improvement |
| Changing from $z$-basis to $x$-basis | 3.1, 3.3, 3.4, 3.5 | 7 \| Major improvement |

**Retention and further learning after post-test solutions were made available**

For the second in-person class, additional data were available to judge students' retention of the concepts that they had learned over the course of the CQS. Students took a midterm exam about three weeks after the post-test which included three questions that asked about concepts covered in questions 2, 6, and 7. (In this writing, corresponding questions between the post-test and the midterm exam are given the same numbers.) Questions 2 and 7 were isomorphic to those asked on the pre- and post-tests, while question 6 was in the context of quantum measurement (which the students had learned in the three weeks between the post-test and midterm exam), which nonetheless required doing a change of basis identical to question 6 on the pre- and post-test. These questions are reproduced in Appendix C. Table 6 shows that when tested on concepts from questions 2 and 6, performance has not changed appreciably, and indeed that for question 7, performance has improved nearly to full correctness.

As seen in Table 6, nearly all students for question 7 were able to demonstrate fluency with a method of changing basis. However, this question dealt with changing between the $y$- and $z$-bases, which do not exhibit a symmetrical relationship the way the $x$- and $z$-bases do. Since this subtlety was not emphasized in the CQS or the original pre- and post-test questions, those students who proceeded as though the relationship was symmetrical were not penalized.

Question 6 was framed as a question about quantum measurement, in which students were asked about the probability of measuring one of the outcomes in the $z$-basis for a state written in the $y$-basis. Again, as discussed above, some students treated the question as though the state was given in the $z$-basis and the measurement basis was the $y$-basis, which is the opposite direction from the one intended and would have yielded an answer of ½ (the correct answer was $\frac{49}{50}$). Students who showed a correct procedure to calculate the incorrect answer were given the benefit of the doubt as to whether they read the given state correctly. Our rubric also avoided penalizing students for thinking that the $|\pm y\rangle$ states are symmetrical with respect to the $|\pm z\rangle$ states (the way the $|\pm x\rangle$ and $|\pm z\rangle$ states are). For example, one student explicitly wrote "This [changing basis from $|\pm z\rangle \to |\pm y\rangle$] is the same as changing basis from $|\pm y\rangle \to |\pm z\rangle$." Thus, students who showed their work received full credit for either answer; no students showed their work to arrive at any other answers. A few students wrote an answer of ½ without showing their work, which could earn them only partial credit. In these cases, it was unclear whether this answer came from the calculation or by incorrectly assuming that a measurement of any observable whose corresponding operator's eigenbasis is not the given basis would always result in either outcome with a probability of ½ (e.g., a system in a state written in the *y*-basis upon which a measurement of the *x*- or *z*-component of spin is made).

**Table 6.** Student performance on similar questions given on a midterm exam about three weeks after the post-test, for in-person class 2 ($N = 27$).

| Question # | Post-test mean | Midterm mean | Normalized gain | Effect size |
|---|---|---|---|---|
| 2 | 96% | 98% | 0.50 | 0.16 |
| 6 | 87% | 83% | - | - |
| 7 | 78% | 94% | 0.75 | 0.55 |

**Summary and conclusion**

We find that, after traditional lecture-based instruction, students perform well on most of the basics of two-state systems such as bra states and inner products, which is encouraging, and struggle substantially more with outer products and approaches to changing basis. We find that, for those latter concepts, a research-validated CQS is highly effective at helping students across all three years. Even though students are performing very well on content covered by questions 1-3 after lecture-based instruction, we recommend that instructors still ask the clicker questions related to these concepts. The questions are written to build on each other, so these would serve as a good warm-up to get students ready for the later questions, without taking very much time.

Moreover, we find that it was not uncommon for students to have difficulty with the rules of matrix multiplication relevant for quantum mechanics, so students' linear algebra preparation cannot necessarily be taken for granted. It was also common for students to add up the matrix elements in an outer product to get a scalar, indicating a conflation with the inner product that is rather understandable, given the visual similarities of the two in Dirac notation. In general, students navigated these concepts substantially better after engaging with the CQS. Interestingly, traditional instruction does not appear particularly successful at teaching the outer product, as many students were confused about this concept on the pre-test.

The most difficult concepts covered by the CQS involved changing basis, which were evaluated by questions 6 and 7 on the pre- and post-tests. Common mistakes included simply dividing the state by $\sqrt{2}$ or answering that the expansion coefficients are the same in both bases. On the post-test, however, the vast majority of students correctly expressed the given state in the desired basis.

In summary, a CQS on the basics and change of basis of two-state systems can be very helpful in reducing the prevalent student difficulties, independent of an online or in-person learning environment, or instructors' individual choices in structuring the course. In all cases, students exhibited moderate to excellent improvement on concepts that they had previously struggled with, including the nature of outer products in matrix representation and procedures of changing basis.

**Ethical statement**

**Acknowledgments**

We thank the NSF for awards PHY-1806691 and PHY-2309260. We thank all students whose data were analyzed and Dr. Robert P. Devaty for his constructive feedback on the manuscript.

**Appendix A**

Note: Minor changes were made to the CQS between years to streamline the concepts presented and eliminate redundancies. Reproduced below is the final version used in the second in-person class.

Notes to Instructor: Basics of 2-State Systems & Change of Basis
- This sequence is meant to familiarize students with the basics as well as the mechanics of changing bases when considering quantum states in two-dimensional Hilbert spaces, in particular spin-1/2 systems.
- Questions 1.1-1.4 present an overview of how a basis can be represented in Dirac notation and using matrix representation. They can be considered a warm-up for students.
- Questions 2.1-2.6 introduce bras and kets in Dirac notation, matrix representation, and inner/outer products expressed in Dirac notation and matrix representation.
- Questions 3.1-3.2 present the transformation of a state $|\chi\rangle$ from the $|\pm z\rangle$ basis to the $|\pm x\rangle$ basis.
- Questions 3.3-3.5 use spectral decomposition of the identity operator $\hat{I}$ as another approach to accomplish the basis transformation.
- Whenever a state is written in matrix representation as opposed to Dirac notation, the $\doteq$ (which stands for "is represented by") is substituted by an equals sign, $=$. However, it should be emphasized to the students that the matrix representation is valid only in the chosen basis (e.g., the basis consisting of eigenstates of $\hat{S}_z$).
- If you are familiar with the states in the notation $|\uparrow\rangle_z$ & $|\downarrow\rangle_z$, $|\uparrow\rangle_x$ & $|\downarrow\rangle_x$, and $|\uparrow\rangle_y$ & $|\downarrow\rangle_y$, they are the same as the states $|+z\rangle$ & $|-z\rangle$, $|+x\rangle$ & $|-x\rangle$, and $|+y\rangle$ & $|-y\rangle$, respectively in the notation used here, such that:

- $\hat{S}_z|z\rangle = \frac{\hbar}{2}|+z\rangle$  $\qquad\qquad \hat{S}_z|-z\rangle = -\frac{\hbar}{2}|-z\rangle$
- $\hat{S}_x|x\rangle = \frac{\hbar}{2}|+x\rangle$  $\qquad\qquad \hat{S}_x|-x\rangle = -\frac{\hbar}{2}|-x\rangle$
- $\hat{S}_y|y\rangle = \frac{\hbar}{2}|+y\rangle$  $\qquad\qquad \hat{S}_y|-y\rangle = -\frac{\hbar}{2}|-y\rangle$
- When $|+z\rangle = \begin{pmatrix}1\\0\end{pmatrix}$, $|-z\rangle = \begin{pmatrix}0\\1\end{pmatrix}$, this means that the basis consists of $\{|+z\rangle, |-z\rangle\}$.

## CQS 1.1
Choosing basis states for a vector space is equivalent to choosing…
    A. **A coordinate system**
    B. Operators
    C. Eigenvalues
    D. A Hilbert space
    E. All of the above

## CQS 1.2
Consider the following statements about a 2-dimensional vector space in which the state of a quantum system lies. Choose all of the statements that are true:
    I. Once you have chosen a basis, you can represent any ket state as a 2 × 1 column matrix.
    II. For every physical observable, there is a corresponding Hermitian operator that can act on states in the vector space.
    III. Once you have chosen a basis, you can represent any operator as a 2 × 2 matrix.

A. I and II only
B. I and III only
C. II and III only
**D. All of the above**
E. None of the above

## CQS 1.3
Choose all of the following statements that are true about the Hamiltonian $\hat{H}_0 = c\,\hat{S}_z$ (where $c$ is a constant) for a spin-1/2 system:
    I. If we choose two different bases (coordinates), $\hat{H}_0$ may be a diagonal matrix in one basis but not in the other.
    II. If the basis states are eigenstates of $\hat{H}_0$, then $\hat{H}_0$ will be a diagonal matrix.
    III. No matter what basis we choose, $\hat{H}_0$ must always be a diagonal matrix by definition.

A. I only
B. II only
**C. I and II only**
D. II and III only
E. None of the above

## CQS 1.4
Choose all of the following statements that are true:

  I.  $|+z\rangle$ and $|-z\rangle$ are always the eigenstates of any given Hamiltonian.
  II.  $|+z\rangle$ and $|-z\rangle$ are the eigenstates of the operators $\hat{S}^2$ and $\hat{S}_z$.
  III.  $|+z\rangle$ and $|-z\rangle$ are the eigenstates of the operators $\hat{S}^2$ and $\hat{\vec{S}}$, where $\hat{\vec{S}} = \hat{S}_x \hat{\mathbf{i}} + \hat{S}_y \hat{\mathbf{j}} + \hat{S}_z \hat{\mathbf{k}}$ in three spatial dimensions.

A. I only
**B. II only**
C. III only
D. I and II only
E. All of the above

## CQS 2.1
Given that $|+x\rangle = \frac{1}{\sqrt{2}}(|+z\rangle + |-z\rangle)$ and $|-x\rangle = \frac{1}{\sqrt{2}}(|+z\rangle - |-z\rangle)$, choose all of the following statements that are true:

  I.  $|+z\rangle = \frac{1}{\sqrt{2}}(|+x\rangle + |-x\rangle)$
  II.  $|-z\rangle = -\frac{1}{\sqrt{2}}(|+x\rangle + |-x\rangle)$
  III.  $|-z\rangle = \frac{1}{\sqrt{2}}(|+x\rangle - |-x\rangle)$

A. I only   B. II only   C. III only   D. I and II only   **E. I and III only**

## CQS 2.2
Given that $|+x\rangle = \frac{1}{\sqrt{2}}(|+z\rangle + |-z\rangle)$, $|-x\rangle = \frac{1}{\sqrt{2}}(|+z\rangle - |-z\rangle)$, $|+y\rangle = \frac{1}{\sqrt{2}}(|+z\rangle + i|-z\rangle)$, and $|-y\rangle = \frac{1}{\sqrt{2}}(|+z\rangle - i|-z\rangle)$, choose all of the following statements that are true:

  I.  $\langle +z|+z\rangle = 1$    and    $\langle -x|+x\rangle = 0$
  II.  $\langle +y|+y\rangle = i$    and    $\langle +z|+y\rangle = \frac{1}{\sqrt{2}}$
  III.  $\langle +x|+x\rangle = 1$    and    $\langle +x|-y\rangle = \frac{1}{2}(1 - i)$

A. I only     B. I and II only     **C. I and III only**
D. All of the above   E. None of the above

## CQS 2.3
Given that $|+x\rangle = \frac{1}{\sqrt{2}}(|+z\rangle + |-z\rangle)$ and $|-x\rangle = \frac{1}{\sqrt{2}}(|+z\rangle - |-z\rangle)$, choose all of the following statements that are true:

  I.  $|+x\rangle\langle -x| = 0$ and $|+x\rangle\langle +x| = 1$
  II.  $|+x\rangle\langle -x| = \frac{1}{2}(|+z\rangle\langle +z| - |-z\rangle\langle -z|)$
  III.  $|+x\rangle\langle -x| = \frac{1}{2}(|+z\rangle\langle +z| - |+z\rangle\langle -z| + |-z\rangle\langle +z| - |-z\rangle\langle -z|)$

A. I only   B. II only   **C. III only**   D. All of the above   E. None of the above

## CQS 2.4
Use the following matrix representations

$|+z\rangle = \begin{pmatrix}1\\0\end{pmatrix}$ and $|-z\rangle = \begin{pmatrix}0\\1\end{pmatrix}$,

$|+x\rangle = \frac{1}{\sqrt{2}}\begin{pmatrix}1\\1\end{pmatrix}$ and $|-y\rangle = \frac{1}{\sqrt{2}}\begin{pmatrix}1\\-i\end{pmatrix}$.

Choose all of the following statements that are true:

I. $\langle +x| = \frac{1}{\sqrt{2}}\begin{pmatrix}1\\1\end{pmatrix}$ and $\langle -y| = \frac{1}{\sqrt{2}}\begin{pmatrix}1\\i\end{pmatrix}$

II. $\langle +x| = \frac{1}{\sqrt{2}}(1\ \ 1)$ and $\langle -y| = \frac{1}{\sqrt{2}}(1\ \ -i)$

III. $\langle +x| = \frac{1}{\sqrt{2}}(1\ \ 1)$ and $\langle -y| = \frac{1}{\sqrt{2}}(1\ \ i)$

A. I only   B. II only   **C. III only**   D. I and III only   E. None of the above

**CQS 2.5**

Use the following matrix representations

$|+z\rangle = \begin{pmatrix}1\\0\end{pmatrix}$ and $|-z\rangle = \begin{pmatrix}0\\1\end{pmatrix}$,

$|+x\rangle = \frac{1}{\sqrt{2}}\begin{pmatrix}1\\1\end{pmatrix}$ and $|-x\rangle = \frac{1}{\sqrt{2}}\begin{pmatrix}1\\-1\end{pmatrix}$.

Choose all of the following statements that are true:

I. $|+x\rangle\langle -x| = \frac{1}{2}\begin{pmatrix}1 & -1\\1 & -1\end{pmatrix}$

II. $|+x\rangle\langle -x| = \frac{1}{2}(1-1+1-1) = 0$

III. $|+z\rangle\langle +z| = \begin{pmatrix}1 & 0\\0 & 1\end{pmatrix}$

**A. I only**   B. II only   C. III only   D. I and III only   E. None of the above

**CQS 2.6**

Use the following matrix representations

$|+z\rangle = \begin{pmatrix}1\\0\end{pmatrix}$ and $|-z\rangle = \begin{pmatrix}0\\1\end{pmatrix}$,

$|+x\rangle = \frac{1}{\sqrt{2}}\begin{pmatrix}1\\1\end{pmatrix}$ and $|-x\rangle = \frac{1}{\sqrt{2}}\begin{pmatrix}1\\-1\end{pmatrix}$,

$|+y\rangle = \frac{1}{\sqrt{2}}\begin{pmatrix}1\\i\end{pmatrix}$ and $|-y\rangle = \frac{1}{\sqrt{2}}\begin{pmatrix}1\\-i\end{pmatrix}$.

Choose all of the following statements that are true:

I. $\langle +x|+x\rangle = \frac{1}{\sqrt{2}}(1\ \ 1)\frac{1}{\sqrt{2}}\begin{pmatrix}1\\1\end{pmatrix} = 1$

II. $\langle +x|+x\rangle = \frac{1}{\sqrt{2}}\begin{pmatrix}1\\1\end{pmatrix}\frac{1}{\sqrt{2}}(1\ \ 1) = \frac{1}{2}\begin{pmatrix}1 & 1\\1 & 1\end{pmatrix}$

III. $|+z\rangle\langle -x| = (1\ \ 0)\frac{1}{\sqrt{2}}\begin{pmatrix}1\\-1\end{pmatrix} = \frac{1}{\sqrt{2}}$

IV. $|+z\rangle\langle -x| = \begin{pmatrix}1\\0\end{pmatrix}\frac{1}{\sqrt{2}}(1\ \ -1) = \frac{1}{\sqrt{2}}\begin{pmatrix}1 & -1\\0 & 0\end{pmatrix}$

A. I only
B. I and III only
**C. I and IV only**
D. II and III only
E. II and IV only

## CQS 3.1
A generic state is written in the $\{|+z\rangle, |-z\rangle\}$ basis as $|\chi\rangle = a|+z\rangle + b|-z\rangle$, but it can be written in another basis. In the $\{|+y\rangle, |-y\rangle\}$ basis, $|\chi\rangle = a'|+y\rangle + b'|-y\rangle$.
Choose all of the following statements that are true:
  I. $a = \langle +z|\chi\rangle$ and $b = \langle -z|\chi\rangle$
  II. $a'$ and $b'$ are the projections of $|\chi\rangle$ along $|+y\rangle$ and $|-y\rangle$, respectively.
  III. $a' = \langle +y|\chi\rangle$ and $b' = \langle -y|\chi\rangle$

A. I only    B. II only    C. I and III only    D. II and III only    **E. All of the above**

## CQS 3.2
A generic state is written in the $\{|+z\rangle, |-z\rangle\}$ basis as $|\chi\rangle = a|+z\rangle + b|-z\rangle$. Given that $|+y\rangle = \frac{1}{\sqrt{2}}(|+z\rangle + i|-z\rangle)$ and $|-y\rangle = \frac{1}{\sqrt{2}}(|+z\rangle - i|-z\rangle)$, choose all of the following statements that are true:

  I. $|+z\rangle = \frac{1}{\sqrt{2}}(|+y\rangle + |-y\rangle)$ and $|-z\rangle = -\frac{i}{\sqrt{2}}(|+y\rangle - |-y\rangle)$
  II. $|\chi\rangle = \frac{a}{\sqrt{2}}|+y\rangle + \frac{ib}{\sqrt{2}}|-y\rangle$
  III. $|\chi\rangle = a\left[\frac{1}{\sqrt{2}}(|+y\rangle + |-y\rangle)\right] + b\left[-\frac{i}{\sqrt{2}}(|+y\rangle - |-y\rangle)\right]$

A. I only    B. II only    C. III only    D. I and II only    **E. I and III only**

Notes to Instructor for CQS 3.2:
- Give students ~2 minutes for CQS 3.2, to calculate what the $|\pm z\rangle$ states are in terms of the $|\pm y\rangle$ states.
- Show that the expansion coefficients in the y-basis $(a'|+y\rangle + b'|-y\rangle)$ simplify to $a' = \frac{a-bi}{\sqrt{2}}$ and $b' = \frac{a+bi}{\sqrt{2}}$, where $a$ and $b$ are the expansion coefficients in the z-basis, $a|+z\rangle + b|-z\rangle$.
- Discuss the special case when $a = \frac{1}{\sqrt{2}}$ and $b = \pm\frac{i}{\sqrt{2}}$. These yield the eigenstates of the y-basis, so that of $a'$ and $b'$, one would be 0 and the other would be 1.

## CQS 3.3
One way to accomplish the process of writing a state in a basis is by acting upon it with the operator $\hat{I}$ written in that basis. Choose all of the following that are valid statements:
  I. $\hat{I} = |+x\rangle\langle +x|$
  II. $\hat{I} = |+x\rangle\langle +x| + |-x\rangle\langle -x|$
  III. $\hat{I} = \begin{pmatrix} 1 & 0 \\ 0 & 1 \end{pmatrix}$ in any basis chosen
  IV. $\hat{I}|\chi\rangle = |\chi\rangle$

A. I and III only    B. III and IV only    C. I, III, and IV only
**D. II, III, and IV only**    E. All of the above

Notes to Instructor for CQS 3.3

- Discuss that the identity operator is the sum of the outer products of a complete set of orthonormal basis states.
- Remind students that the identity operator $\hat{I}$ acting on any state returns that same state. $\hat{I}$ can be placed anywhere based upon convenience, and is analogous to multiplying a scalar by 1.
- Show (or have students confirm) that the sum of the outer products for a complete set of orthonormal basis vectors returns the identity operator in matrix form in any basis.

**CQS 3.4**
Given that the identity operator $\hat{I}$ multiplied by any state returns that state, choose all of the following that are an equivalent way of writing the state $|\chi\rangle$:

I. $|\chi\rangle = \hat{I}|\chi\rangle = (|+x\rangle\langle+x| + |-x\rangle\langle-x|)\,|\chi\rangle$
II. $|\chi\rangle = \hat{I}|\chi\rangle = (|+y\rangle\langle+y| + |-y\rangle\langle-y|)\,|\chi\rangle$
III. $|\chi\rangle = \hat{I}|\chi\rangle = (|+x\rangle\langle+x| + |-y\rangle\langle-y|)\,|\chi\rangle$

A. I only  B. II only  C. III only  **D. I and II only**  E. All of the above

**CQS 3.5**
If we want to express our state in the x basis, we can write the following:
$$|\chi\rangle = \hat{I}|\chi\rangle = (|+x\rangle\langle+x| + |-x\rangle\langle-x|)\,|\chi\rangle$$
Choose all of the following that correctly represent the state $|\chi\rangle = a|+z\rangle + b|-z\rangle$ in the $\{|+x\rangle, |-x\rangle\}$ basis:

I. $|\chi\rangle = (|+x\rangle\langle+x| + |-x\rangle\langle-x|)\,(a|+z\rangle + b|-z\rangle)$
II. $|\chi\rangle = |+x\rangle\langle+x|a|+z\rangle + |-x\rangle\langle-x|b|-z\rangle$
III. $|\chi\rangle = a|+x\rangle + b|-x\rangle$

**A. I only**  B. II only  C. III only  D. All of the above  E. None of the above

**Appendix B**

The post-test questions are reproduced below. Pre-test questions are isomorphic with small changes, such as different given states. Students were given the following information:
- Whenever a state is written in matrix form in a given basis as opposed to Dirac notation, the symbol $\doteq$ (which stands for "is represented by") is substituted by an equals sign, $=$, for convenience. For example, $|+z\rangle \doteq \begin{pmatrix}1\\0\end{pmatrix}$ will be written as $|+z\rangle = \begin{pmatrix}1\\0\end{pmatrix}$.
- If you are familiar with the states in the notation $|\uparrow\rangle_z$ & $|\downarrow\rangle_z$, $|\uparrow\rangle_x$ & $|\downarrow\rangle_x$, and $|\uparrow\rangle_y$ & $|\downarrow\rangle_y$, they are the same as the states $|z\rangle$ & $|-z\rangle$, $|x\rangle$ & $|-x\rangle$, and $|y\rangle$ & $|-y\rangle$, respectively, such that:

$$\hat{S}_z|+z\rangle = \frac{\hbar}{2}|+z\rangle \qquad \hat{S}_z|-z\rangle = -\frac{\hbar}{2}|-z\rangle$$
$$\hat{S}_x|+x\rangle = \frac{\hbar}{2}|+x\rangle \qquad \hat{S}_x|-x\rangle = -\frac{\hbar}{2}|-x\rangle$$
$$\hat{S}_y|+y\rangle = \frac{\hbar}{2}|+y\rangle \qquad \hat{S}_y|-y\rangle = -\frac{\hbar}{2}|-y\rangle$$

In the z-basis, the basis states $|\pm z\rangle, |\pm x\rangle$ and $|\pm y\rangle$ can be obtained by diagonalizing $\hat{S}_z, \hat{S}_x$, and $\hat{S}_y$, respectively, where:

$$\hat{S}_z = \frac{\hbar}{2}\begin{pmatrix} 1 & 0 \\ 0 & -1 \end{pmatrix}$$

$$\hat{S}_x = \frac{\hbar}{2}\begin{pmatrix} 0 & 1 \\ 1 & 0 \end{pmatrix}$$

$$\hat{S}_y = \frac{\hbar}{2}\begin{pmatrix} 0 & -i \\ i & 0 \end{pmatrix}$$

- $\langle \phi | \chi \rangle = \langle \chi | \phi \rangle^*$ for any two generic states $|\chi\rangle$ and $|\phi\rangle$, where * denotes the complex conjugate: $(a \pm ib)^* = (a \mp ib)$
- The eigenstates of $\hat{S}_x$ in terms of the eigenstates of $\hat{S}_z$ are as follows:

$$|+x\rangle = \frac{1}{\sqrt{2}}(|+z\rangle + |-z\rangle)$$

$$|-x\rangle = \frac{1}{\sqrt{2}}(|+z\rangle - |-z\rangle)$$

1. Consider the following statements about a 2-dimensional vector space in which the state of a quantum system lies. Are the following statements true or false?

   _____ a) For every physical observable, there is a corresponding Hermitian operator that can act on states in the vector space.

   _____ b) Once you have chosen a basis, you can represent any operator as a $2 \times 2$ matrix.

   _____ c) Once you have chosen a basis, you can represent any ket state as a $2 \times 1$ column matrix.

2. Given that $|+z\rangle = \begin{pmatrix} 1 \\ 0 \end{pmatrix}, |-z\rangle = \begin{pmatrix} 0 \\ 1 \end{pmatrix}$

   Write $|+y\rangle = \frac{1}{\sqrt{2}}(|+z\rangle + i|-z\rangle)$ and $|-x\rangle = \frac{1}{\sqrt{2}}(|+z\rangle - |-z\rangle)$ as matrices in this basis.

3. Given that in the z basis

   $$|\chi\rangle = \frac{1}{\sqrt{34}}\begin{pmatrix} 3i \\ 5 \end{pmatrix} \qquad |\psi\rangle = \frac{1}{\sqrt{17}}\begin{pmatrix} 4 \\ -i \end{pmatrix}$$

   Represent $|\chi\rangle$ and $|\psi\rangle$ in the form $|\chi\rangle = c|+z\rangle + d|-z\rangle, |\psi\rangle = e|+z\rangle + f|-z\rangle$ (that is, find $c, d, e,$ and $f$).

4. Given that in the z-basis, $|\chi\rangle = \frac{1}{\sqrt{34}}\begin{pmatrix} 3i \\ 5 \end{pmatrix}$ and $|\psi\rangle = \frac{1}{\sqrt{17}}\begin{pmatrix} 4 \\ -i \end{pmatrix}$, find $\langle \chi | \psi \rangle$. Show or explain your work to get credit.

5. Given that in the z-basis, $|\chi\rangle = \frac{1}{\sqrt{34}}\begin{pmatrix} 3i \\ 5 \end{pmatrix}$ and $|\psi\rangle = \frac{1}{\sqrt{17}}\begin{pmatrix} 4 \\ -i \end{pmatrix}$, find $|\psi\rangle\langle\chi|$. Show or explain your work to get credit.

6. Write the state $|\chi\rangle = \frac{3}{\sqrt{10}}|+x\rangle + \frac{1}{\sqrt{10}}|-x\rangle$ in the form $|\chi\rangle = a'|+z\rangle + b'|-z\rangle$ (that is, find $a'$ and $b'$). Show or explain your work.

7. Write the state $|\chi\rangle = \frac{3}{\sqrt{10}}|+z\rangle + \frac{1}{\sqrt{10}}|-z\rangle$ in the form $|\chi\rangle = a'|+x\rangle + b'|-x\rangle$ (that is, find $a'$ and $b'$). Show or explain your work.

Note that we start here in the z-basis and wish to convert to x-basis, the opposite of the preceding question.

## Appendix C

The questions on the midterm that corresponded to selected questions on the pre- and post-test are reproduced below.

2. Given that $|+z\rangle = \begin{pmatrix} 1 \\ 0 \end{pmatrix}$, $|-z\rangle = \begin{pmatrix} 0 \\ 1 \end{pmatrix}$

Write $|+y\rangle = \frac{1}{\sqrt{2}}(|+z\rangle + i|-z\rangle)$ and $|-y\rangle = \frac{1}{\sqrt{2}}(|+z\rangle - i|-z\rangle)$ as matrices in this basis.

6. Consider a state $|\chi\rangle = \frac{3}{5}|+y\rangle + \frac{4}{5}|-y\rangle$. If you measure $S_z$, what is the probability of obtaining a value $\frac{\hbar}{2}$?

7. Write the state $|\chi\rangle = \frac{3}{\sqrt{10}}|+z\rangle + \frac{1}{\sqrt{10}}|-z\rangle$ in the form $|\chi\rangle = a'|+y\rangle + b'|-y\rangle$ (that is, find $a'$ and $b'$). Show or explain your work.